\documentclass[final,5p,twocolumn]{elsarticle}

\usepackage{graphicx}
\usepackage{amssymb}
\usepackage{lineno}
\usepackage{amsmath}
\usepackage{algorithm2e}
\journal{Robotics and Autonomous Systems}

\begin{document}

\begin{frontmatter}


\title{An Ontology to Support Collective Intelligence in  Decentralised Multi-Robot Systems}



\author{Pragna Das$^{1}$, Vincent Hilaire$^{2}$ and Llu{\'\i}s Ribas-Xirgo$^{3}$}

\address{$^{1}$Pragna Das is a Researcher in Microelectronics and Electronic Systems Department,
        Universitat Aut\`{o}noma de Barcelona (UAB), Campus UAB, 08193 Bellaterra, Spain\\
        pragna.das@uab.cat\\
        $^{2}$Vincent Hilaire is responsible for Multi-Agent Systems team of the SeT Laboratory, 
        Univ Bourgogne Franche-Comté, UTBM, 
        IRTES-SET EA 7274/IMSI F-90010,
        Belfort cedex, France\\
        vincent.hilaire@utbm.fr\\
        $^{3}$Llu{\'\i}s Ribas-Xirgo is with the Faculty of Microelectronics and Electronic Systems Department, Universitat Aut\`{o}noma de Barcelona (UAB), 
        Campus UAB, 08193 Bellaterra, Spain\\
        Lluis.Ribas@uab.cat\\
        }

\begin{abstract}
In most multi-robot systems, conditions of the floor, battery and mechanical parts contribute to the costs incurred in performances of movements and tasks. The time to complete performance is dependent on all these factors and thus reflects the costs incurred. The relation between performance times and these factors are not directly derivable, though, performance time has a direct correlation with discharge of batteries. When movement is a performance, travel time of an edge is the performance time. When travel times can be estimated to obtain close-to-real values, they become different than heuristics costs and depict the real states which are impossible to obtain from heuristics. This facilitates path planning algorithms to choose the edges with least real travel times or costs to form the path. Nevertheless, a good estimation is dependent on historical data which are close in time. But, there are situations when all the travel times for one or more edge(s) are not available for the entire duration of operation of the MRS to an individual robot. Then, it is imperative for that robot to gather the necessary travel times from others in the system as a reference observation. This work involves devising a mechanism of information sharing between one robot to others in the system in a form of a common ontology-based knowledge. With the help of this ontology, travel time is obtained by any robot, whenever necessary, to obtain accurate estimates for itself. These obtained travel times are traveling experiences of other robots in the system. Still, they can be used to estimate travel time in that robot as model of travel time has an exploration factor which depicts the change of travel time in that robot. This model uses others' travel time as observation and self-exploration factor to estimate future travel time. This greatly helps the MR to estimate travel times more accurately and precisely. The accurate estimation affects route planning to be more precise with reduced cost. The total cost of paths obtained using travel times estimated through sharing is 40\% less on average than that of paths generated through travel times without sharing. 
\end{abstract}

\begin{keyword}
ontology, multi-agent, behavior-based system, multi-robot system, decentralized multi-robot system, decentralized control, control system, collective intelligence


\end{keyword}

\end{frontmatter}

%\linenumbers

\section{Introduction}
\label{intro}
%Many advantages in the multi-robot systems (MRS) for manufacturing and logistics have ushered due to projects in Industrie 4.0, 
%One of the most crucial aspects of a multi-robot systems (MRS) is the organizing of flow of decisions. The objective of control of MRS is to accomplish two classes of tasks, system level tasks like flocking, task assignment and mitigating collision, while maintaining communication among all the robots \cite{ICRA2016}. 
%Planning in a centralized organization of control \cite{FarinelliMRS} of a multi-robot systems (MRS) is relatively easy than a distributed control as the former comprises of a leader (agent) who is responsible for planning and assigning to other agents (robots), who just obey the direction. But, maintainability and communication can be threatened with a slight damage in the leader agent in centralized control for MRS. Planning and communication can be made robust in a decentralized organization of control where dedicated control are provided for each robot and communication is made possible among all the robots. In this organization, communication and planning in MRS is not compromised as there is no leader \cite{} and the system can still operate, removing the damaged agent. Thus, decentralized flow of control is suitable to the current demands of MRS like adaptability, decentralization, real-time operation and control, service orientation, modularity \cite{grauIECON}. 
The organization control in MRS for manufacturing and logistics is mostly centralized in the state of the art with few exceptions like \cite{dec1, slidingDec, panRobotics, FarinelliMRS} where the architecture is decentralized, as it is relatively easy than a distributed control. However, the decentralized architecture in \cite{panRobotics} is focused on only the planning of route to generate feasible, sub-optimal and collision-free paths for multiple MRs. Thus, the system architecture is not general enough to handle different kinds of control and planning functions. In \cite{slidingDec}, the linear dynamic model is generated for a specific task of collectively transporting load in automated factories. The sliding mode controller is provided through a non-linear terms with bounds. This kind of stochastic control using dynamic model of the robot is useful in simple cases where the controller depends on minimal information which is available to the robot, unaware of the dynamics of the environment. 

On the other hand, there are various investigations conducted using partially observable Markov decision processes (POMDPs) to solve the general decentralised control and planning problems in MRS \cite{auctionPOMDP, mitPOMDP}, due to the the improvement of the general concepts of multi-agent systems (MAS). However, these solutions are computationally expensive and provide sub-optimal solution. Also, the requirements of scalability and robustness in a smart factory is not met with these solutions. Likewise, the problems solved in latter case using POMDP ignore the aspect of improving the cooperative functions based on performances or state of individual MRs and their environment. 

Thus, the problem still persists where a decentralized system architecture is necessary for MRS which will be scalable and robust, yet computationally inexpensive. 

The robot computational architecture in an MRS is classified into 4 categories, such as, deliberative, reactive, hybrid and behavior-based \cite{Michaud2016}. Though, all the four categories of control have their advantages and limitations, each contribute with interesting but different insights and not a single approach is ideal \cite{Michaud2016}. Nevertheless, current demands of smart factories are adaptability, real-time response and modularity which is served excellently by behavior-based control \cite{Michaud2016}. Moreover, decentralized control can be efficiently designed, implemented and handled through behavior-based systems. Numerous investigations have been carried out recently towards solving the problems of formation control in MRS using behavior-based systems \cite{Lee2018}, while much attention is not paid toward behavior-based control of MRS in logistics and transportation tasks. There are few investigations where this direction is investigated \cite{icraBehavior}, \cite{gonzalez2016behavioral}, but these behavior-based MRSs lack the much needed involvement of support and help by other MRs to each MR in decision making. This help can be successful when cost incurred by one robot for a particular performance can be shared to other robots to estimate their costs to carry out the same performance. These costs arise through the different battery and floor conditions while performances. Thus, the time to complete performance is dependent on all these factors and thus reflects the costs incurred. When movement is a performance, travel time of an edge is the performance time. When travel times can be estimated to obtain close-to-real values, they become different than heuristics costs and depict the real states which are impossible to obtain from heuristics. Nevertheless, a good estimation is dependent on historical data which are close in time. But, there are situations when all the travel times for one or more edge(s) are not available for the entire duration of operation of the MRS to an individual robot. Then, it is imperative for that robot to gather the necessary travel times from others in the system as a reference observation. We demonstrate this concept with the following example. 

\begin{figure}[h]
\centering\includegraphics[width=1\linewidth]{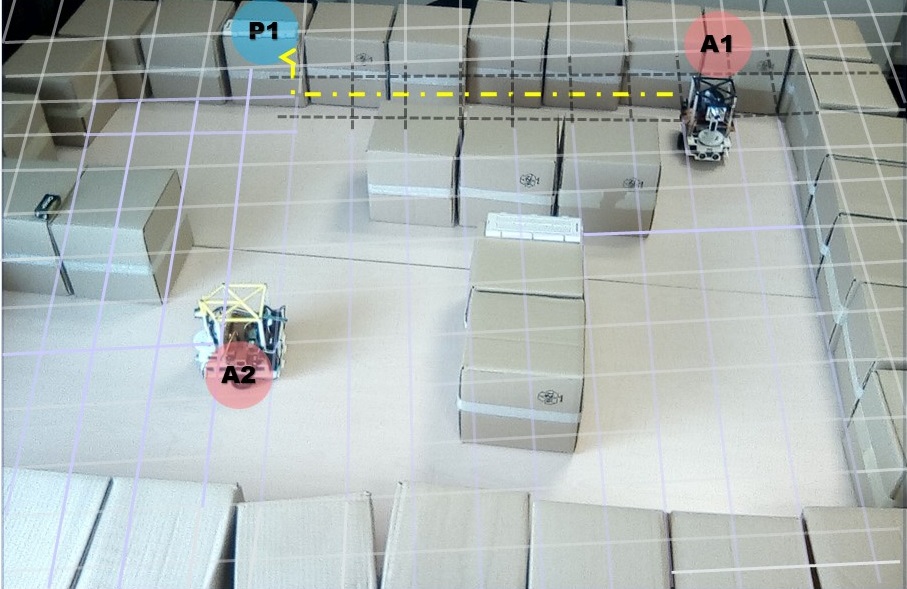}
\caption{Problem: An Example Scenario}
\label{figExmplCollaborative}
\end{figure}
%\begin{figure}[th]
 %     \centering
  %    \framebox{\parbox{3.3in}{%We suggest that you use a text box to insert a graphic (which is ideally a 300 dpi TIFF or EPS file, with all fonts embedded) because, in an document, this method is somewhat more stable than directly inserting a picture.
   %   \includegraphics[scale = 0.35]{examplePic.jpg}
%}}
      
 %     \caption{Problem: An Example Scenario}
  %    \label{figExmplCollaborative}
   %\end{figure}
Figure~\ref{figExmplCollaborative} illustrates a scaled down MRs-based internal transportation system in a factory. Traversing a path is considered as a task in this example. Let, at any instance of time, $t_0$, $A$1 is assigned to carry some material to $P$1 through the computed path marked by the dotted line. Again, at time $t_m$ ($m$ > 0), $A$1 needs to carry same material to $P$1. But at $t_m$, $A$1 will need more time and energy to reach $P$1 than at $t_0$ due to mainly two reasons. First, the battery capability of $A$1 has decreased due to execution of previous tasks. Also, the condition of the part of the floor, designated by the given path, can get  deteriorated (as marked by black dotted lines). As said previously, travel times of edges depict states of battery and floor condition \citet{wafPragna}. So, the travel times at previous instances are useful to estimate travel time at $t_m$, if only battery state has changed. But, condition of floor has also changed. This can only be anticipated through travel times at $t_m$ if the robot has traversed that part of the floor in the previous or nearly previous time instance. Nevertheless, travel time from other MRs who has traversed that part in nearly previous instance can be useful to $A$1, along with its own travel times at previous instances to estimate its travel time at current. Thus, travel times of these two sources are useful to estimate it's future travel time. This work addresses this area of investigation where each MR get information like travel time of edges from other MRs in order to make better decisions. %In this way,  decisions for planning for each robot can be improved.

The above example explains that the amount of time and energy required to complete a task has an existing correlation with state of charge of batteries and environmental conditions. These time and energy can be formed as cost coefficients to express state of battery's charge and environment. These cost coefficients can be of various forms like travel time of edges, rotating time, loading time, \textit{et~cetera} depending on the functions. 
Further, they can serve as a deciding factor in several planning decisions for better cost efficient decisions. 
However, these cost coefficients need to be either known apriori or estimated to be used in decision making. In case of knowing apriori, observations of these costs in various forms like travel time of edges, rotating time, \textit{et~cetera} need to be measured for all possibilities, which is not only cumbersome but also impractical. Hence, estimating them during run-time is a good solution. But, estimation requires observation of the same at previous instances. The observations values can be gathered from the beginning of first decision making and can be used in subsequent calls for estimation. The first few iterations of decision making is a learning phase to gather few observations to start the estimation. But, an MR may need to estimate the travel time of one or more edge which it did not traverse previously. This can be mitigated by sharing the observation value from other MR who has travelled that edge in nearly previous instance. This way the knowledge sharing can help an MR to estimate the travel cost for an unexplored edge at current instance. In the example, $A$1 has travelled the edges in the region (marked by dotted line) towards $P$1 long back at $t_0$ and hence it does not have the latest information about its condition at $t_m$. In this case, travel time of edges, annotated with time stamps, along the region marked by dotted line from other MRs who has travelled it in nearly previous instance, must be communicated to $A$1, so that $A$1 can utilize it while estimating it's own cost at $t_m$. 

The travel times have inherent contexts like time stamp and the edge between pair of nodes. This underlying context has been exploited to  form a semantic knowledge sharing mechanism to communicate the costs of edges inform of travel times in this work. This is instrumental in deriving more accurate estimates of travel times to ascertain cost at current time in each MR. This improves decisions in each MR for efficiency where MRs help each other to gather states of environment and other factors. Moreover, all the MRs are autonomous and have their own control separately, which make the whole system decentralized. This type of control is implemented using a behavior-based system, to utilize the benefits of both decentralized architecture and behavior-based system. 

The subsequent sections elaborate on the background (Section~\ref{bckgrnd}), problem statement and  contribution (Section~\ref{probContr}), methodology (Section~\ref{methControl} and Section~\ref{semantics}) and implementation (Section~\ref{implControl} and Section~\ref{ontology}). Results of utilising the proposed methodology is tabulated and analyzed on Section~\ref{res}, while discussions and conclusions are put forward in
%\section{Background}
\label{bckgrnd}
\section{Problem statement and contribution}
\label{probContr}
This work addresses the problem of building a decentralized system  architecture where the planning decisions can be based on the dynamically changing state of MRs and their environment. This paves the way towards robustness and scalability in the MRS. One of the most suitable methods of control for MRS is behavior-based system as it can handle significant dynamic changes with fast response and enforce adaptability, few of the major requirements of current smart factories \cite{Michaud2016}. The system architecture for the MRS in current work is developed using concepts of behavior-based system with specific behaviors for planing and task execution.

Current work implements an MRS for automated logistics where each robot need to transport materials to designated placeholders or racks, termed as ports. 
Also, \textbf{reaching a particular port} by an MR is considered as a task, along with route computing being considered as a decision making process. Thus an MR is required to traverse from one node to another in a floor, described by a topological map. This enables the MRs to perform single task at a time.

\begin{figure}[h]
\centering\includegraphics[scale = 0.38]{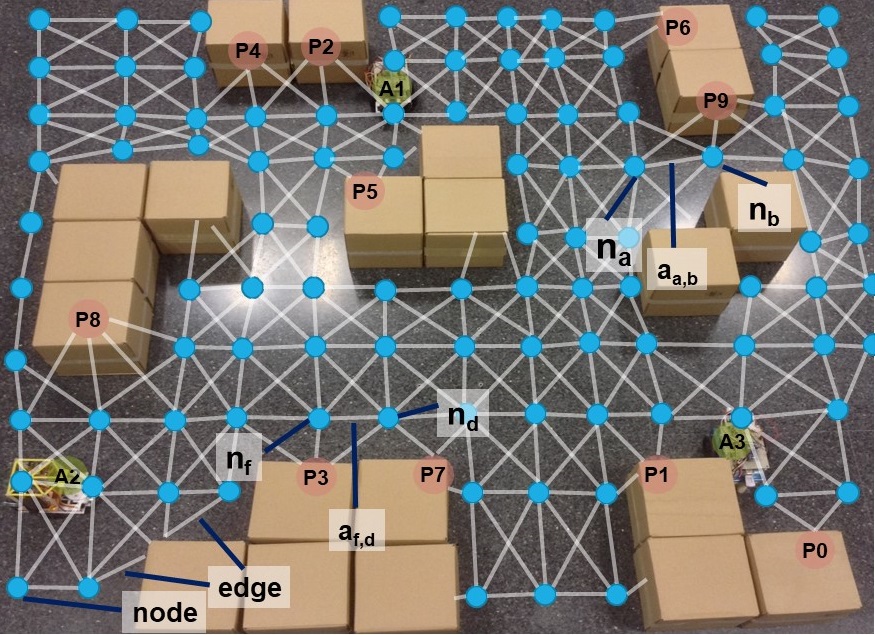}
\caption{Problem description}
\label{map}
\end{figure}

The travel time of each arc (like $a_{a,b}$, $a_{f,d}$) in a floor map given in Figure~\ref{map} is influenced by energy exhaustion, condition of floors, physical parameters of robot, among others, which incurs cost. Thus time to traverse an arc by an MR or \textit{travel time} can be conceptualized as its cost coefficients. In this work, \textit{travel time} is considered as weight or cost for an edge. This is formalized as $X_{p,q}(k)$ to denote travel cost from $n_p$ to $n_q$, where $k$ is the instance of time of traversing an edge. $X_{p,q}(k)$ is time-varying from the perspective that at a particular instance of the time, the cost of that particular arc is dependent on battery discharge and condition of the floor which changes over passage of time. A path $P$ is formed as a series of arcs between connecting nodes for an MR and thus $P$ can be defined as 
\begin{equation}
\label{pathL}
P = \langle a_{a,b}, a_{b,e}, a_{e,g}, a_{g,j}, ...........\rangle
\end{equation}

Now, the cost of traversing $P$ can be written in a form $C_P$ of 
%where the edge cost of traversing from one node to another is expressed. Thus, $P$ can be written
\begin{equation}
\label{pathC}
C_P = \langle X_{a,b}(k), X_{b,e}(k), X_{e,g}(k),...,... ...........\rangle
\end{equation}
The elements of $C_P$ are required to be identified for each call of path planning. From now on, $X_{p,q}(k)$ will be written as $X(k)$ for simplicity. 
For continuous performance of the MR, path needs to be computed for the MR after it reaches a destination. Let, at $i$th call of path planning, path cost was
\begin{equation}
\label{pathCI}
C^i_P = \langle X^i_{a,b},..., X^i_{e,g}, ....., X^i_{q,r}, ..... \rangle
\end{equation}
Now, in any instance, an MR may need to traverse an arc which it had traversed in previous instances. Let at $j$th ($j$\textgreater$i$+1) call of path planning, estimation of $X(k)$ for $a_{e,g}$ is required. As , the MR do not have the observation of $X(k)$ for $a_{e,g}$ at the previous instance. It can only use the $X(k)$ for $a_{e,g}$ obtained during traversing $P$ after $i$th call. In this scenario, the obtained estimate can be significantly inaccurate which has the potential to produce inaccurately optimized path. Thus, the observation of $X(k)$ for $a_{e,g}$ at $j$th call can be also obtained from other robots performing in the system which has traversed that arc in the previous instance or in a nearly previous past instance. Moreover, at $j$th call, estimation of $X(k)$ for some edge may be needed which that MR has not been yet traversed during all its previous traversals. This can be also solved by fetching the observation data of $X(k)$ for the required edge from other robots which have traversed that at previous or nearly previous instances. 

Improved estimated values of $X(k)$ can be obtained by transferring the right knowledge from one robot to another, which will generate more cost efficient decision. Thus, an information sharing framework is imperative to be formed in order to improve the estimation and in turn improve the decisions where robots can support and help each other in their decisions.  

The contribution of this work include the following
\begin{itemize}
    \item A completely decentralised system architecture is developed based on behavior-based system in a hierarchical model for each robot, which ensures scalability and improves robustness
    %\item The \textbf{utility} based factors, \textbf{reaching a particular destination} is considered as a task and route computing is considered as a decision making for control of movements
    \item A semantic knowledge sharing mechanism is devised in each robot to share estimated values of travel times of one robot to others. This eventually helps in obtaining better estimates of travel times in each robot, which produces for more optimal path with minimum path costs.
\end{itemize}

\section{Behavior-based decentralised multi-robot system}
\label{controller}
\subsection{Methodology}
\label{methControl}
\begin{figure}[h]
\centering\includegraphics[scale = 0.38]{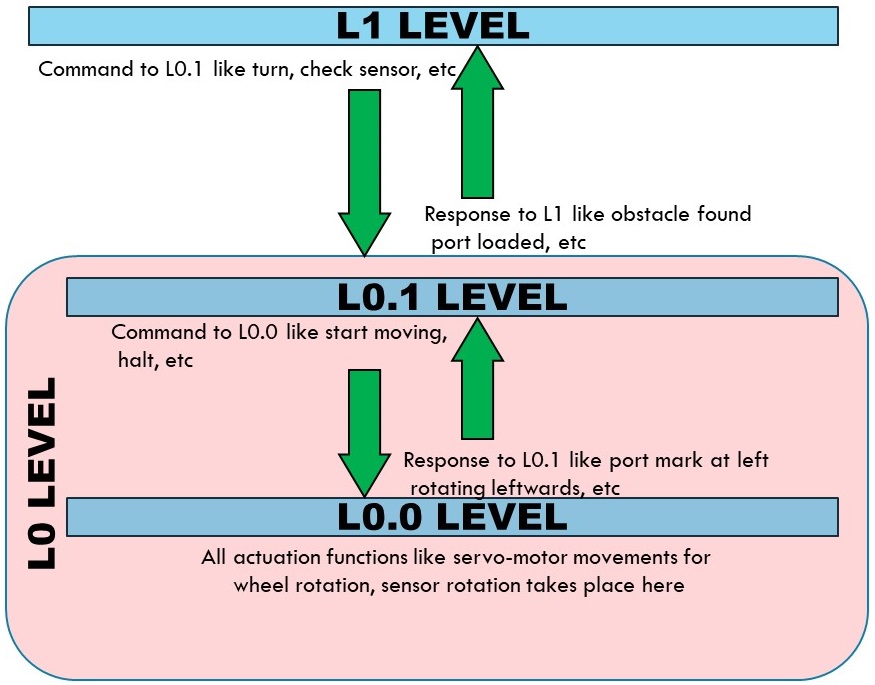}
\caption{Controller architecture}
\label{agvControl}
\end{figure}

The objective of this work is to form an MRS with a decentralized flow of control, suitable to logistics. The flow of control is based on the concept of sub-sumption, where each robot has the same sub-sumption model. Each robot is capable of taking the decision itself, with the capability of gathering information about the environment from other MRs. This sub-sumption model achieves the goal making each MR autonomous.   
The sub-sumption model involves the control structure to be organized in layers one above the other with increasing level of competencies and each level can interact with all other levels with messages. This technique of flow of control is described in Figure~\ref{agvControl}, which consists of two major control layers. 

The top most layer is $L$1 level and the $L$0 level is below it. The $L$0 level is divided into two sub levels $L$0.1 and $L$0.0 levels respectively. The $L$1 level is the agent level control layer where it functions on all the agents in the transportation system and is engaged in controlling more complex functions like finding path, organizing task, finding destination poses, \textit{et~cetera} for each of the robots. The $L$0 level functions on each of the agents individually and controls the movements. Each robot has its own $L$0.0 and $L$0.1 levels respectively which controls the movements in each of them. Here, the $L$0.0 level communicates with the $L$0.1 level and have no communication with the $L$1 level. The $L$0.1 level is the intermediate level which communicates both with $L$0.0 level and $L$1 level. The control levels functions in co-ordination with each other to control the movements of the robots in the environment \cite{Norouzi2013}. Thus, essentially the MRs in the system are autonomous. Moreover, the top most $L$1 level is responsible for intelligent decision making for task assignments and path traversal, based on the available \textit{travel time}, which represents knowledge about the individual robot and the environment. 
\subsection{Implementation}
\label{implControl}
\begin{figure}[h]
\centering\includegraphics[scale = 0.33]{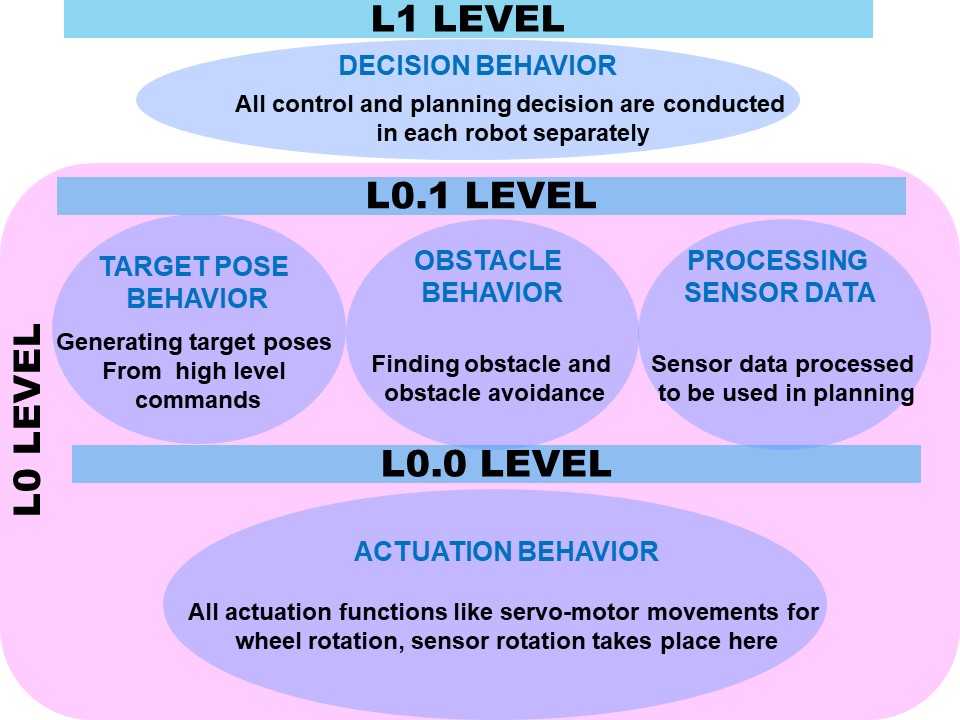}
\caption{Behaviors for the layers of control}
\label{masLayers}
\end{figure}

The control technique \cite{Ismael2015},\cite{Ismael2017}, described in previous section, is achieved using behaviors as the building block of both decision-making level ($L$1 level) and action execution level ($L$0 level). Separate sets of behaviors are designed for two layers as illustrated in Figure~\ref{masLayers}, which is based on the control framework proposed by R. Brooks in \cite{brookControl86}. The hierarchical control framework has three behavioral levels and each level has an objective and a corresponding output, formed as commands or replies. Moreover, the process of execution of all levels start simultaneously. However, the output in the form of commands from the highest level ($L$1) need to pass on to the next priority level ($L$0.1) for it to start execution, and similar process is followed in $L$0.0 level. This happens because of the hierarchical framework and the command from the highest-priority behavioral level is required as input to process the low-priority behavioral level. On the other hand, the replies from the lower level act as a feed back to the  control rules of the higher level which determines the final decision and output of the control framework.

The general design of an MR is considered in the prototype system which consists of servo-motors to rotate wheels and camera. The sensing is conducted with infra-red sensors and camera. The beagle-bone forms the processor for the robot. Each MR in the system is autonomous with its own three level of behavioral control framework. The following are the behaviors developed in each level.

\begin{itemize}
    \item In $L$0.0 level, actuation behaviors are developed. This behavior conducts the starting of motors for wheel rotation, camera movements and infra-red sensor movements. The commands which refer to target poses are obtained from $L$0.1 level in this behavior using extended finite state machines to conduct the movement of the individual robots. The sensor readings are transferred to $L$0.1 for processing as feedback of commands. 
    \item In $L$0.1 level, three behaviors are developed. They are generating target poses from high level commands like destination port, finding obstacle and obstacle avoidance, processing sensor data to be used in planning and decision in $L$1. All these behaviors are developed using extended finite state machine. 
    
    \item In $L$1 level, decision making behaviors are developed which are finding paths and assigning tasks. These behaviors are developed using extended finite stacked-state machine. Also, behavior of maintaining and sharing the knowledge of \textit{travel time} is developed. More details about the sharing mechanism is provided in next sections. 
\end{itemize}

All these three levels of behavior correspond to a singular behavior for a single robot and this is repeated for each MR. Thus, the control flow in the MRS is decentralized. The knowledge sharing mechanism is incorporated in the behavior of $L$ so that each robot can communicate through them. The highest level ($L$1 level) is implemented on the desktop computer in our model to reduce communication costs among the $L$1 level agents in each MR. The next two lower levels are implemented on body on individual MR using embedded system techniques. The decisions for planning need the information about the states of itself and environment. As behavior in $L$ level conducts the process of decision, the knowledge sharing process is realized in $L$ in each level which provides each robot an opportunity to seek help about states of environment from other MRs.

\begin{figure}[h]
\centering\includegraphics[scale = 0.6]{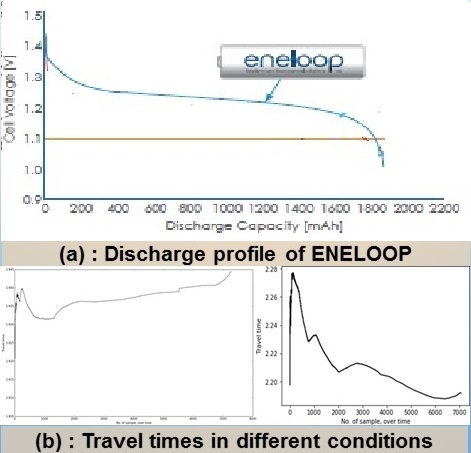}
\caption{change pic}
\label{disvstt}
\end{figure}

As discussed in Section~\ref{intro}, travel time $X$($k$) (Section~\ref{probContr}) for a particular edge provides the necessary representation of state of robots and environment. A direct correlation has been found between $X$($k$)  with state of charge of batteries and conditions of floor in the prototype system. This is depicted in Figure~\ref{disvstt}. Part (a) plots the cell voltage of Li-ion batteries over time, Part (b) plots the progressive mean of observed values of $X$($k$) for $m$th edge with the change of state of charge of batteries and Part (c) plots the observed value of $X$($k$) for the same edge with both the change of state of the charge of batteries and the floor condition. The floor is changed from rough at the beginning to smooth during the experiment. 

The plot (b) shows that progressive mean of $X$($k$) increase first, then steadily decrease and then increase gradually till complete discharge. Thus values of $X$($k$) first increase due to sudden fall of cell voltage at beginning, then decreasing fast due to cell voltage increasing fast to a steady level and the values gradually decrease towards complete discharge of batteries. Thus, a correlation between $X$($k$) is observed through plots (a) and (b). On the other hand, the increase in progressive mean of $X$($k$) is longer than that of plot (b) at equal battery capacity. The longer increase of values of $X$($k$) in (c) can be attributed to the rough floor, as more energy is required to traverse in rough surface. Plot of $X$($k$) in different conditions of floor demonstrate that travel time can reflect not only state of charge of batteries \cite{wafPragna} but also environmental conditions. 

During the run-time of MRS, the estimation of $X$($k$) is conducted for all necessary edges while finding the optimal path. Thus, estimated values of $X$($k$) will be generated at every instance of control decisions, producing a pool of estimated values. More significantly, every estimated value of travel time has inherent context associated with it, which when shared with the other MRs help in the estimation of $X$($k$) in them. This concept is elaborated in the next section.

\section{Information sharing in behavior-based control}
\label{infoShare}
\subsection{Semantics in travel time}
\label{semTT}
An MRS is dynamic as its states change over time. Also, it is evolving as it gathers more and more knowledge about its states through the course of its operation. Moreover, the source of knowledge of an MRS is distributed to each of its constituent robots. The behavior in $L$1 has the role of finding paths using Dijkstra's algorithm. Dijkstra's algorithm needs to know the estimated $X$($k$) for the concerned edges to decide the path as $X$($k$) designates the cost of traveling the edge. Now, there are possibilities when an MR has not yet traversed many edges. The estimation of $X$($k$) for these edges depends on the obtained travel cost of them from other MRs. Thus, knowledge sharing mechanism improves the estimation of $X$($k$) for accuracy. This will be instrumental to Dijkstra's algorithm to produce better optimal paths with minimum path costs.

The $X$($k$)s originate from each MR depending on the instance of travelling, zone of the floor, previously traversed edge, neighboring AGVs, state of charge, \textit{et~cetera}. All these factors provide context to the estimated values of $X$($k$). 

\begin{figure}[h]
\centering\includegraphics[scale = 0.36]{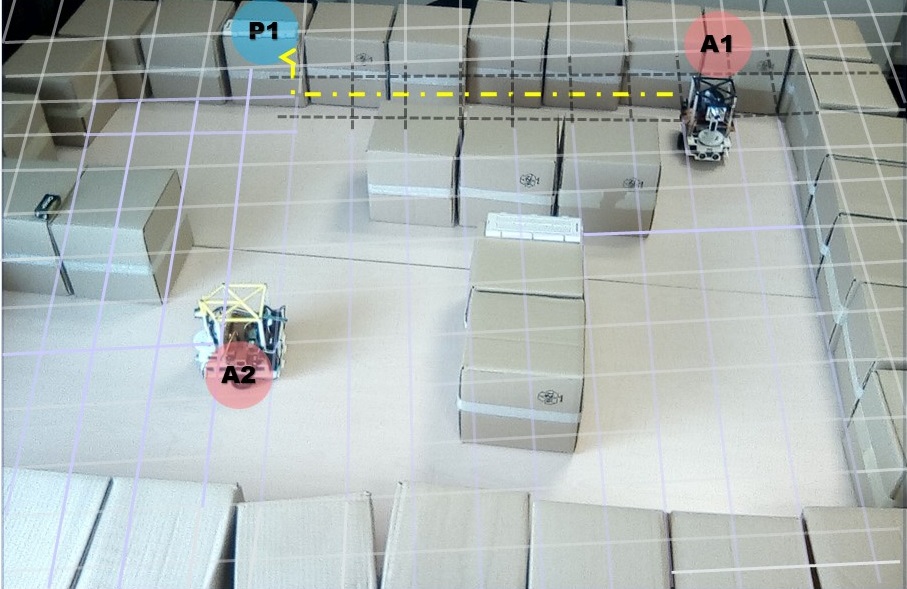}
\caption{change pic}
\label{explSemtc}
\end{figure}

For example, in Figure~\ref{explSemtc}, the $X$($k$) of an edge by $A$1 at $t_m$ will be different than that at $t_0$ due to discharging of batteries as explained in Figure~\ref{disvstt}. On the other hand, $X$($k$) for $n$th edge ($n$ $\neq$ $m$) by $A$2 in a different zone (marked by double dotted line) will be different than that of $m$th edge, though both $m$th and $n$th edge can have same lenghth. This happens due to different states of floor. Moreover,  $X$($k$) for $n$th edge by $A$1 will be different than that by $A$2 at any $t_i$ because of differently discharged batteries for different previous tasks.
Thus, estimated travel time provides contextual information representing state of charge, condition of floor, instance of travelling. 

These values of $X$($k$) at a particular instance for a particular edge of one MR provide contextual information about cost for that edge to other MRs when communicated. Hence, semantics can be built from these knowledge of travel time as they have inherent contextual information. They convey information about the costs of traversing through different edges in the topological map, which describes the factory floor.
\subsection{Using semantics for knowledge sharing}
\label{semantics}
\begin{figure}[h]
      \centering
      \framebox{\parbox{3.3in}{%We suggest that you use a text box to insert a graphic (which is ideally a 300 dpi TIFF or EPS file, with all fonts embedded) because, in an document, this method is somewhat more stable than directly inserting a picture.
      \includegraphics[scale = 0.47]{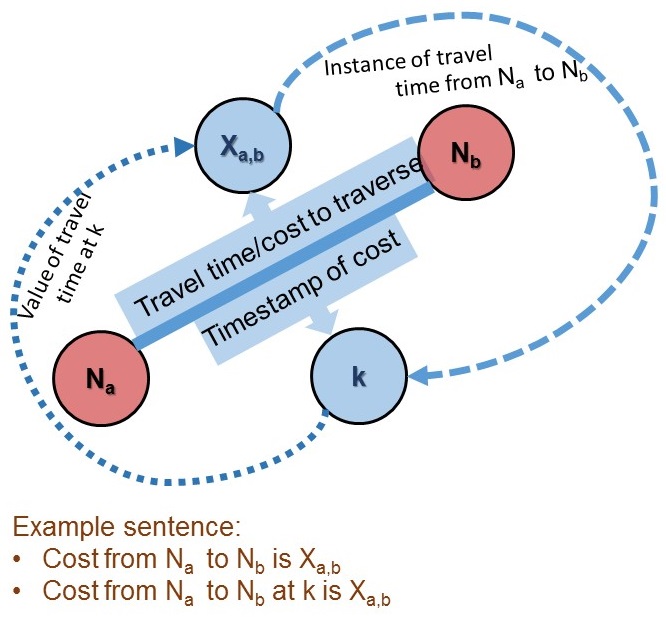}
}}
      
      \caption{An example of semantic relationship in MRS}
      \label{semPic}
   \end{figure}

Semantic relationships are built in this work to form semantic sentences in order to fetch the values of travel times with the inherent contextual information. This concept is illustrated in Figure~\ref{semPic}. 

From the above example, a semantic sentence can be derived as
\begin{itemize}
\label{egSentence}
\item Cost from node $N_a$ to node $N_b$ is $X_{a,b}$ at time instance $k$ 
\end{itemize}
where, $k$ is the instance of estimation. $N_a$ and $N_b$ refer to specific nodes, travel time $X_{a,b}$ refer to specific kind of cost. Cost refer to specific kind of utility expenditure while performing the task. Thus, \textbf{cost} establishes the relationship between nodes $N_a$ and $N_b$ and \textbf{travel time} $X_{a,b}$. When the system knows the meaning of \textbf{nodes}, \textbf{utility cost}, \textbf{travel time}, then the above sentence will convey some meaning to the system. This is precisely the method of developing semantics in the MRS in order to convey the contextual meaning instilled in travel time to the $L$ level controller.
\subsection{Ontology to represent semantics}
\label{ontForSem}
The most traditional, flexible and useful method of representing knowledge using semantics is expressions based on subject, predicate and object logic \cite{SegaranPTSW}. Positioning the obtained knowledge is the next progressive step which is defined in philosophical terms as ontology.
Ontology helps to create order and define relationships among things useful to an application.
A domain specific ontology is developed in this work to efficiently store, access and communicate meaningful semantics across all the MRs in the system regarding the real-time travel costs of edges.

There are significant advantages of implementing ontology for the already mentioned application of this work. 
\begin{itemize}
\label{dbComparison}
\item \textbf{Conceptualization of information}: 
An ontology is defined explicitly to form a specification for a shared conceptualization of a pool of knowledge \cite{sageOnto}, \cite{ontoGRUBER1993}, \cite{ontoSTUDER1998}. Ontologies define the concepts of the domain formally and explicitly making further modifications or reversals less cumbersome. 
\item \textbf{Data representation}: 
Ontology is based on dynamic data representation where a new instance definition is not constrained to a definite rule. Thus adding new elements is easy and fast as and when required. This virtue of ontology is essentially beneficial to share the knowledge of travel time in MRS. The number of travel time grows with the increase of operation time. 
Moreover, reasoners in ontology solve the problem of data parity, integrity and adhering to constraints. When a new element is added to an ontology, the reasoner performs to check the integrity of the information. This capability of ontology makes the knowledge sharing method in the MRS flexible yet robust. Data addition in MRS is not required to be done on all instances and when it is added the reasoner checks for data integrity and new information can be added smoothly without adhering to rules, previously defined.  
\item \textbf{Modeling technique}: Ontology possesses the capability to express semantic concepts. In case of MRS, conveying the contextual information inherent to any cost parameter like travel time requires this semantic expressiveness than just defining or extracting data. Moreover, the pool of knowledge gathered in the MRS through travel time or similar parameters need to be reused which is only possible through the descriptive logic models of ontology.
\end{itemize}

In nutshell, ontology provide an unrestricted framework to represent a machine readable reality, which assumes that information can be explicitly defined, shared, reused or distributed. Moreover, information can also be interchanged and used to make deductions or queries. Such representation is imperative for representing the travel time for reasons described above.
\subsection{Application of ontology}
\label{ontology}
Semantics is an efficient way to communicate enough meaning which can actuate some action. 
The focus on representing semantic data is through entities. Semantic models are property oriented and semantic entities are members of a class. Semantic classes are defined on properties, it is also possible to define classes in terms of value of a property. A property type is \textit{object property} when it signifies some abstract property like character, contribution, virtue \textit{et~cetera}. A property type is a \textit{data property} when it signifies some literal value. On the other hand, classes can have any of the type of properties. The subclasses are defined which can avail all the properties of the superclass. The properties have range and domain. Range is the source type of a property, while domain is the destination type of the property. 

\begin{figure}[ht]
      \centering
      \framebox{\parbox{3.3in}{%We suggest that you use a text box to insert a graphic (which is ideally a 300 dpi TIFF or EPS file, with all fonts embedded) because, in an document, this method is somewhat more stable than directly inserting a picture.
     \includegraphics[scale = 0.2]{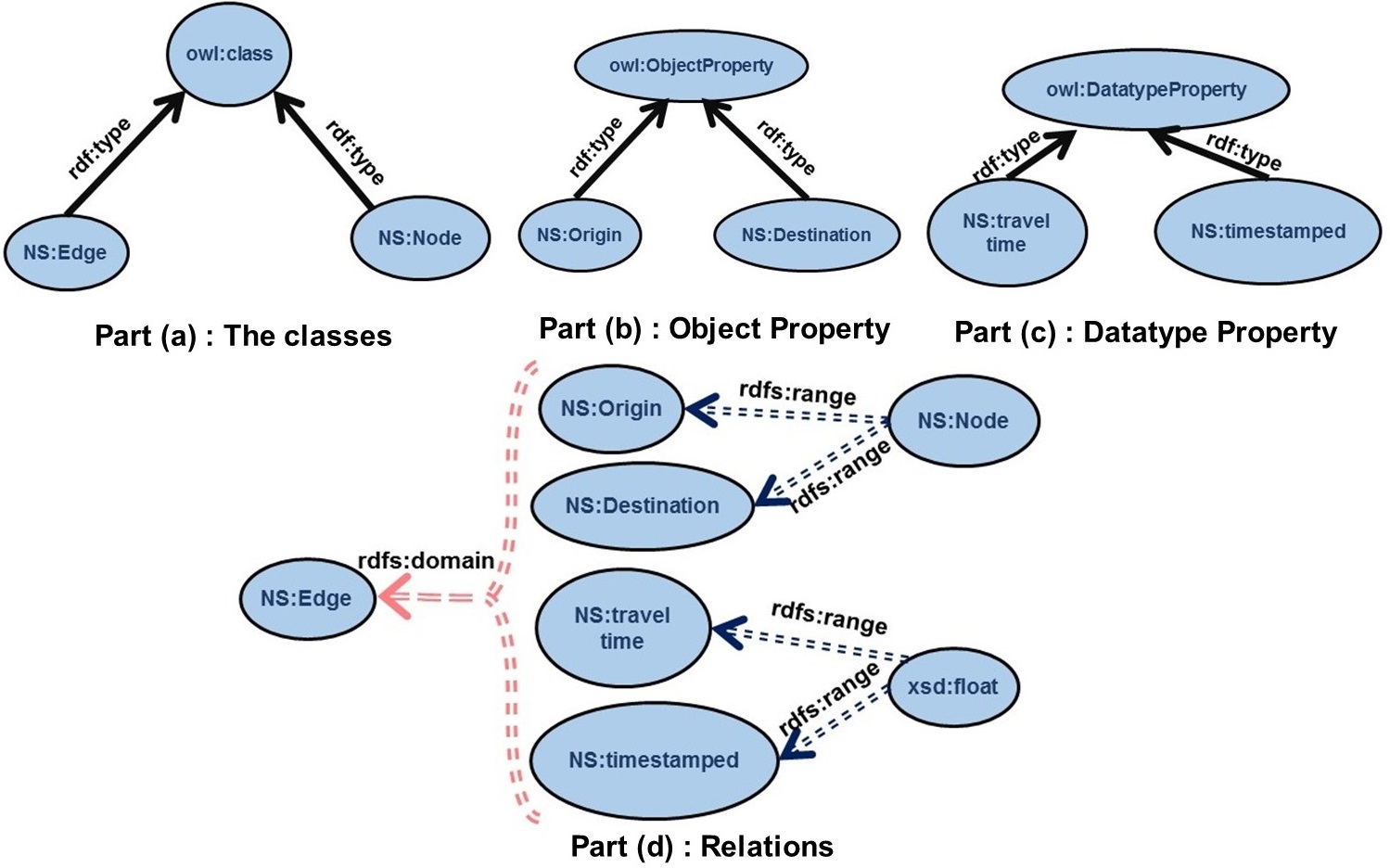}
}}
      
      \caption{Ontology}
      \label{desOnto}
   \end{figure}
   
Based on these concepts, the ontology stores and shares the knowledge of travel time (Figure~\ref{desOnto}). The ontology has two types of classes (\textbf{owl:Class}), \textbf{NS:Edge} and \textbf{NS:Node}, as shown in Figure~\ref{desOnto}. Thus, \textbf{NS:Edge} and \textbf{NS:Node} are subclasses of \textbf{owl:Class}. 
There are two properties a class can possess, \textbf{owl:ObjectProperty} and \textbf{owl:DatatypeProperty}. \textbf{NS:Origin} and \textbf{NS:Destination} are of types of \textbf{owl:ObjectProperty}, while  \textbf{NS:tt}, \textbf{timeStamped} are of types of \textbf{owl:DatatypeProperty}.

The range of \textbf{NS:Origin} is subclass \textbf{NS:Node}, being the source type of a property, while domain is \textbf{NS:Edge} being the destination type of the property. Similarly, the range of \textbf{NS:Destination} is subclass \textbf{NS:Node}, being the source type of a property, while domain is \textbf{NS:Edge} being the destination type of the property. On the other hand, the range of \textbf{NS:tt} is a float, being the source type of a property, while domain is \textbf{NS:Edge} being the destination type of the property. Similar is the case for \textbf{timeStamped}. The tupled relationships are formed by using these domain and range connections. For example, let $m$th edge be between nodes $n_g$ and $n_h$. $X$($k$) for $m$th edge at $k$ can be formed as \textbf{NS:tt} value at \textbf{timeStamped} value $k$ for the $m$th individual of subclass \textbf{NS:Edge} whose \textbf{NS:Origin} is individual $n_g$ of subclass \textbf{NS:Node} and \textbf{NS:Destination} is individual $n_h$ of subclass \textbf{NS:Node}.  This semantic sentence can be disintegrated into several subject, predicate and object logic to derive the necessary $X$($k$). For example,  
\begin{itemize}
    \item individual $m$th edge is of type \textbf{NS:Edge}
    \item individual $n_g$ is of type \textbf{NS:Node} 
    \item individual $n_h$ is of type \textbf{NS:Node}
    \item $m$th edge has \textbf{NS:Origin} $n_g$
    \item $m$th edge has \textbf{NS:Destination} $n_h$
    \item $m$th edge has \textbf{NS:tt} $X$($k$)
    \item $m$th edge has \textbf{timeStamped} $k$
\end{itemize}

This way the \textbf{owl:ObjectProperty} and \textbf{owl:DatatypeProperty} of the subclass \textbf{NS:Edge} provides the $X$($k$) for the $m$th edge. Also, the $X$($k$) gets a context about its edge (between a pair of nodes) and time stamp. 
The advantage of this ontology lies in this formation, as discussed in previous Section~\ref{dbComparison}, where any new element can be inserted through these property formations without being restrained semantically. With the use of ontology, travel time $X$($k$) can be efficiently stored annotated with a pair of nodes demarcating the edge and the time stamp of traversing it.

The structure illustrated in Figure~\ref{desOnto} shows the formation of ontology which is replicated in each robot in the MRS. 
Thus, when the information of travel cost for any edge for any time instance is required by any MR, $X$($k$) for that edge at the required time stamp can be retrieved from ontology of other MRs. This shared information from other MRs can provide as observation or historical data for those edges which either have not been yet travelled or have been travelled long back. This helps in achieving accurate estimates of $X$($k$) of these edges. 

For example, in Figure~\ref{explSemtc}, when $A$2 requires to estimate $X$($k$) for edges through the marked zone (marked by dotted line), the historical observation data of $X$($k$) in that zone can be obtained from the ontology of $A$1 whi h as traversed those edges in previous or nearly previous instance. The estimated values at current instance become more accurate using $X$($k$) of the same edges by $A$1 at previous instances.

This information can be sought by the $L$1 level behaviors in any MR to other $L$1 level behaviors in other MRs. Thus, this ontology fulfills the mechanism of knowledge sharing inside the $L$ level behaviors. A co-operative approach in achieved through this knowledge sharing for better cost efficient decisions in each MR, which in turn enhances the cost efficiency of the MRS. 
\section{Retrieval of travel time and using in estimation}
The sharing of travel time to all MRs is implemented through ontology in each of them to generate better estimate of travel time among all (Section~\ref{need2share}). This section describes the methodology of using travel time of others in the estimation process of an MR. 

The travel time of an MR is modelled using bi-linear state dependent time series \cite{priestley1988}, which is described in Section~\ref{exp2} in Chapter~\ref{costPathPln}. This is again produced here for convenience. 
The bi-linear model, provided in equation~\ref{bilinear}, is used
to model the change of travel costs depending upon all the
previous travel costs.\\ 
\begin{align}
\label{bilinear}
&X(k)+a_1X(k-1)+.....+a_jX(k-j)\\ \nonumber
&=\xi_k+b_1\xi(k-1)+...+b_l\xi(k-l)\\ \nonumber
&+\sum\sum c_{rz}\xi(k-r)X(k-z) \nonumber
\end{align}
The model described in equation~\ref{bilinear} is a special case of the general class of non-linear models called state dependent model (SDM) \cite{priestley1988}. In equation~\ref{bilinear}, $X$($k$) denotes the edge travel cost at $k$ and $\xi$ at $k$ denotes the inherent variation of the edge travel cost. In equation~\ref{bilinear}, $X$($k$) depends on all the previous values of $X$ and $\xi$, whose number is provided by the variables $j$ and $l$. However, a fixed number of previous values of $X$ and $\xi$ is used for estimation of current $X$ like an window which moves with increase of time. This fixed size of this window is termed as \textit{regression number} and it is chosen as a design parameter, designated by $j$ and $l$. The double summation factor over $X$ and $\xi$ in equation~\ref{bilinear} provides
the nonlinear variation of $X$ due to state of batteries and
changes in environment.

The state space form of the bi-linear model is given in equation~\ref{bStateEqn} and equation~\ref{bObsEqn}.
\begin{align}
    \label{bStateEqn}
%\begin{aligned}
&s(k) = F(s(k-1))s(k-1) + V\xi_k+G\omega_{k-1}\\
\label{bObsEqn}
&Y(k) = Hs(k-1) + \xi_k+ \eta_k
\end{align}
The equation~\ref{bStateEqn} is the state equation which provides the next state from the current state. 
In equation~\ref{bStateEqn}, the state vector $s$($k$) is of the form 
$(1, \xi_{k-l+1},...., \xi_k, X_{k-j+1},......, X_k)^T$. The state vector contains the edge costs obtained progressively over time from $X_{k-j+1}$ to $X_k$. The variable $\xi$ provides values of innovation or evolution of edge costs over the time as the exploration proceeds. Here, $j$ denote number of previous edge costs to be included in the state vector among all edges included in the path till $k$th instance. Also, $l$ denotes the number of previous evolution values of these edges. The $\xi$ values are specific for each MR and originate from the changes in travel time of the particular MR. The values of $\xi$ are obtained by sampling using the observation data of travel time. This observation data is obtained for the static online estimation of travel time (Section~\ref{exp1} in Chapter~\ref{costPathPln}). The $\xi$ values obtained through this method represents the projection of change of travel time. 
Though, these sampling method does not produce the perfect data to represent the change of travel time, this is suitable to this simple case where cost factor of one task is considered. This method should be improved for the case where cost factors of two or more tasks are to be considered. 

The matrices of equation~\ref{bStateEqn} are $F$, $V$ and $G$ which are explained in the following. 
\[
F = 
\begin{bmatrix}
    1\mspace{18.0mu}0\mspace{18.0mu} 0\mspace{18.0mu}\dots\mspace{18.0mu}0\mspace{18.0mu}\vdots\mspace{18.0mu}0\mspace{18.0mu} 0\mspace{18.0mu}\dots\mspace{18.0mu} 0\mspace{18.0mu}0\\
    0\mspace{18.0mu}0\mspace{18.0mu} 1\mspace{18.0mu}\dots\mspace{18.0mu}0\mspace{18.0mu}\vdots\mspace{18.0mu}0\mspace{18.0mu} 0\mspace{18.0mu}\dots\mspace{18.0mu} 0\mspace{18.0mu}0\\
    0\mspace{18.0mu}0\mspace{18.0mu} 0\mspace{18.0mu}\dots\mspace{18.0mu}1\mspace{18.0mu}\vdots\mspace{18.0mu}0\mspace{18.0mu} 0\mspace{18.0mu}\dots\mspace{18.0mu} 0\mspace{18.0mu}0\\
    0\mspace{18.0mu}0\mspace{18.0mu} 0\mspace{18.0mu}\dots\mspace{18.0mu}0\mspace{18.0mu}\vdots\mspace{18.0mu}0\mspace{18.0mu} 0\mspace{18.0mu}\dots\mspace{18.0mu} 0\mspace{18.0mu}0\\
    \vdots\mspace{18.0mu}\vdots\mspace{18.0mu}\vdots\mspace{18.0mu}\dots\mspace{18.0mu} \vdots\mspace{18.0mu}\vdots\mspace{18.0mu}\vdots\mspace{18.0mu} \vdots \mspace{18.0mu}\dots\mspace{18.0mu} \vdots\mspace{18.0mu} \vdots\\
    0\mspace{18.0mu}0\mspace{18.0mu} 0\mspace{18.0mu}\dots\mspace{18.0mu}0\mspace{18.0mu}\vdots\mspace{18.0mu}0\mspace{18.0mu} 1\mspace{18.0mu}\dots\mspace{18.0mu} 0\mspace{18.0mu}0\\
    0\mspace{18.0mu}0\mspace{18.0mu} 0\mspace{18.0mu}\dots\mspace{18.0mu}0\mspace{18.0mu}\vdots\mspace{18.0mu}0\mspace{18.0mu}0 \mspace{18.0mu}1\mspace{18.0mu}\dots\mspace{18.0mu}0\\
    0\mspace{18.0mu}0\mspace{18.0mu} 0\mspace{18.0mu}\dots\mspace{18.0mu}0\mspace{18.0mu}\vdots\mspace{18.0mu}0\mspace{18.0mu} 0\mspace{18.0mu}0\mspace{18.0mu}\dots\mspace{18.0mu} 1\\
    \mu\mspace{18.0mu}\psi_l\mspace{18.0mu}\psi_{l-1}\mspace{18.0mu}\dots\psi_1\vdots-\phi_k-\phi_{k-1}\dots-\phi_1
\end{bmatrix}
\]
The number of rows of $F$ depends on the number of $regression\_no$ and given by (2*$regression\_no$ + 1). The matrix $F$ contains many new terms like $\psi$ , $\phi$ , $\mu$. The $\psi$ terms are denoted as in equation~\ref{psiterms}
\begin{equation}
\label{psiterms}
\begin{aligned}
\psi_l =b_l+ \sum_{i=1}^{l} c_{li}X(k-i)\\
\end{aligned}
\end{equation}
All the $\phi$ terms in $F$ are constants. The term $\mu$ is the average value of $X$ till $k$th instance. Thus, the state transition matrix $F$ depends on the travel times of the previously traversed edges. 
Also, the matrix $V$ is denoted as 
\[
V = 
\begin{bmatrix}
    0 & 0 & 0 & \dots & 1 & \vdots & 0 & 0 & \dots & 1
\end{bmatrix}
\]

The number of rows of $V$ is again given by (2*$regression\_no$ + 1). 
The equation~\ref{bObsEqn} is the observation equation which forms the observation for the current instance. 
The matrix in equation~\ref{bObsEqn} is $H$ which is described as
\[
H = 
\begin{bmatrix}
    0 & 0 & 0 & \dots & 0 & \vdots & 0 & 0 & \dots & 1
\end{bmatrix}
\]

The observation is formed by multiplying the $H$ matrix with the state vector $s$ and added with the innovation at the current instance. In both equation~\ref{bStateEqn} and equation~\ref{bObsEqn}, $s$($k$-1) denotes the state vector at current instance. Here, $s$($k$-1) is of the form $(1, \xi_{(k-1)-l+1},...., \xi_{k-1}, X_{(k-1)-j+1},......, X_{k-1})^T$. The $X$ values in this vector are the travel times obtained for the edges which are already explored and included in the path. But, the available travel times may not be enough to fulfill all the data till the previous instance. Thus, the travel times for the previous instance which are not available are gathered from other MRs. In order to gather this data, the relevant edge costs are queried in the ontology of other MRs. Then after retrieval of the data they are filled in the state vector for both equation~\ref{bStateEqn} and equation~\ref{bObsEqn}. Both the equations have $\xi$ which corresponds to the innovation or change of travel time. Thus, this factor plays the role of projecting the travel time of the particular MR at some particular instance. In equation~\ref{bStateEqn}, this factor contributes not only to the formation of state but also forms the state equation to predict the next state. In equation~\ref{bObsEqn}, this $\xi_k$ is added to the product of $H$ and $s$($k$-1) to form observation. The product of $H$ and $s$($k$-1) is $X_{k-1}$ The addition of $X_{k-1}$, $\xi_k$ and error term $\eta_k$ produces the observation $Y$($k$). In this way, the travel time of other MRs are used in the model for estimation of travel time. The estimation is done by Kalman filtering. The equations obtained after applying Kalman filtering this bi-linear model are explained in Section~\ref{exp2} in Chapter~\ref{costPathPln}. The same process is continued to obtain the travel time of relevant edges. 

This travel times are the instruments to decide the path using Dijkstra's algorithm. This whole process is summarized in Algorithm~\ref{sharingDiijkstra}. 

\begin{algorithm}
 \caption{Dijkstra's algorithm using dynamic estimation of travel time}
 \label{sharingDiijkstra}
%\begin{algorithmic}[1]
    \SetKwInOut{Input}{Input}
    \SetKwInOut{Output}{Output}

    \underline{Initialise\_Single\_Source} $(V,E,s)$\\%\;
    \Input{$V$-list of nodes, $E$-list of edges, $s$-source node}
    \Output{$d$[$v$]-attribute for each each node, $\pi$[$v$]-predecessor of each node}
%\Function{Create\_Topo\_Map}{$A,r,c$}\Comment{A-grid map, r-number of rows, c-number of columns}\\
\SetAlgoLined
%\KwResult{Write here the result }
 %totalrowcount := row number of grid map\;
 %totalcolcount := column number of grid map\;
 \For{each $x_i \in V$}{
    $\pi$[$x_i$] = infinity\\
    $d$[$x_i$] = NIL\\
    }
 $d$[$s$] = 0\\ 
\underline{findedgedCost} $(u,v,j)$\\%\;
    \Input{$u$-current node, $v$- neighbor node}
    \Output{$w$- estimated travel\_time (cost) from $u$ to $v$}
        $findPredEdge$($u$)\\
        $prev_x$ := $x_(prevEdge)$\\
        $w$ = $estimateKF$($prev_x$,$j$,$X$)\\
\underline{findPredEdge} $(u)$\\%\;
    \Input{$u$-current node}
    \Output{$prevEdge$-edge connection $u$ and $predU$}
        $prevEdge$ = edge between $u$ and $predU$\\
\underline{estimateKF} $(prev_x,j,X)$\\%\;
    \Input{$prev_x$-$x_(j-1)$, $j$-instance for estimation, $X$- observation variable}
    \Output{$x_j$-travel cost at current j for current edge}

    Apply Kalman filtering to find $s_j$ and return $x_j$
   
\underline{Relax} $(u,v,w)$\\%\;
    \Input{$u$-current node, $v$- neighbor node, $w$- estimated travel\_time (cost) from $u$ to $v$}
    \Output{$d$[$v$]-attribute for each each node, $\pi$[$v$]-predecessor of each node}
    \If{$d$[$v$] $> d$[$u$] + $w$($u,v$)}{
      $d$[$v$] = $d$[$u$] + $w$($u,v$)\\
      $\pi$[$v$] = $u$\\
    }
\underline{Main} $(V, E, w, s)$\\%\;
    \Input{$V$-list of nodes, $E$-list of edges, $w$-edge weight matrix, $s$-source node}
    \Output{$\pi$[$v$]-predecessor of each node}
    $P$ := NIL\\
    $Q$ := $V$\\
    $j$ := 0\\
    \While{$Q != $0}{
       $j$ = $j$+1
       $u$ := Extract min ($Q$)\\
       $P$ := $P$ $\bigcup u$\\
       \For{each $v \in Adj$[$u$]}{
         $w$ = $findedgedCost$($u$,$v$, $j$)\\
         $relax$($u$,$v$,$w$)\\
       }
    }
%\end{algorithmic}
\end{algorithm}
\section{Experiment and Results}
\label{res}
This work proposes a behavior-based control method which uses online estimated \textit{travel time} as a decision parameter for computing optimal routes between pairs of ports. 
\subsection{Experiment-I: Behavior-based decentralized control system for MRS based logistics} 
\label{exp1}
A prototype multi-robot system for logistics in factory is developed based on the proposed behavior-based decentralized planning and control method. The experimentation platform is briefly described in this section to provide elaborate explanation of the experiment.
A scaled down prototype of automated indoor logistics system is built. 

\begin{figure}[h]
\centering\includegraphics[scale = 0.36]{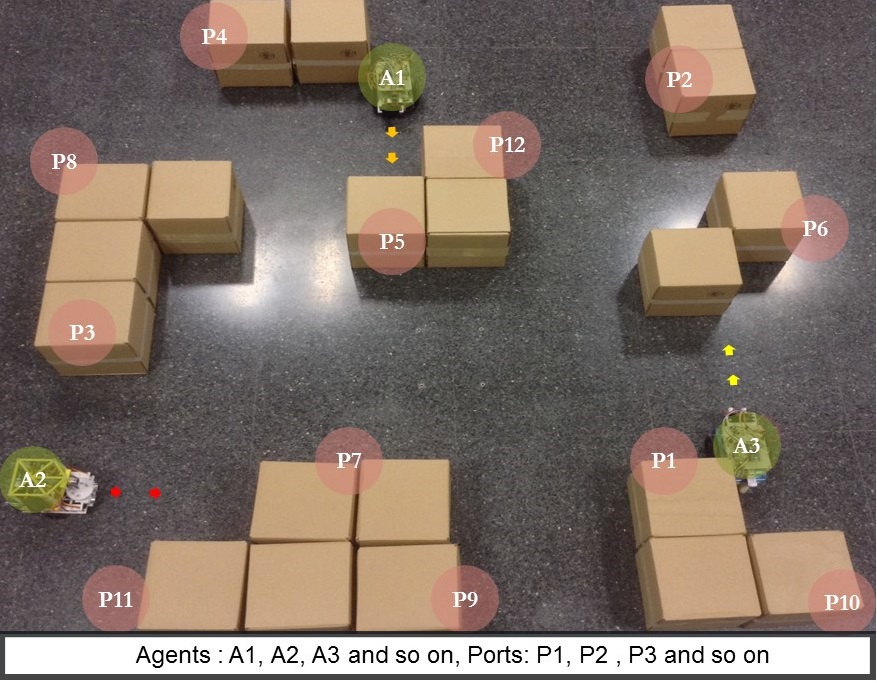}
\caption{Environment of MRS}
\label{envrn}
\end{figure}

\begin{figure*}[t]
\centering\includegraphics[scale = 0.28]{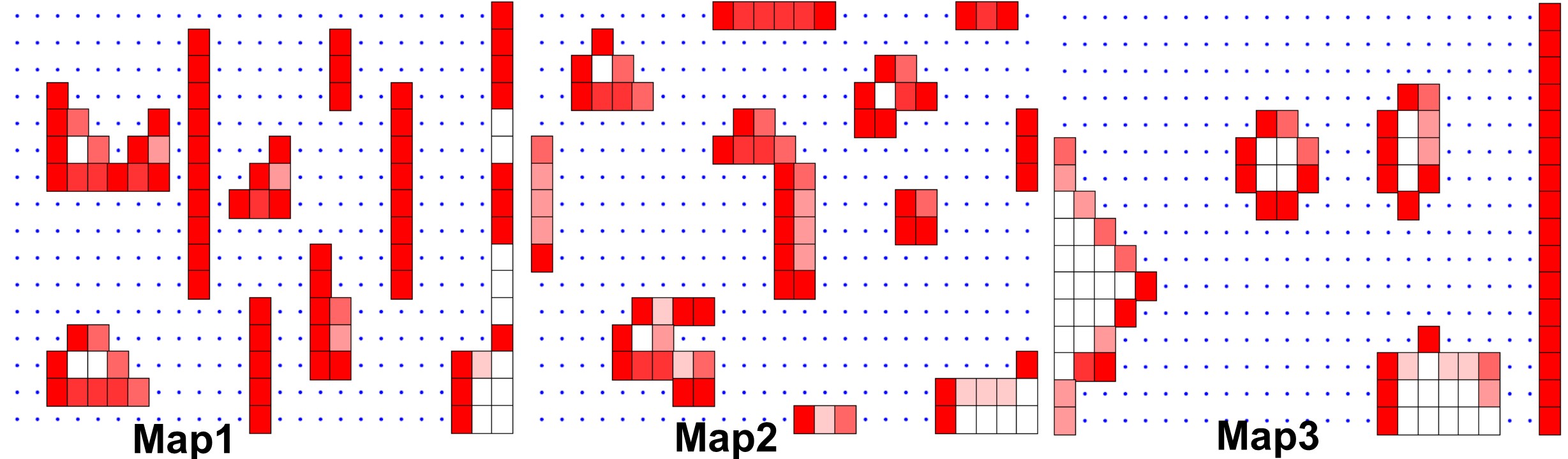}
\caption{The three topology maps}
\label{3maps}
\end{figure*}

An environment has been developed using uniform sized boxes as shown in Figure~\ref{envrn} for the robots to work, doing single task at a time and is named as single-task robot. The boxes create a closed labyrinth path to navigate. Also designated ports are marked on the boxes. The floor is described in three different topological maps. These maps are provided in Figure~\ref{3maps}. 

The control structure is same in each MR which consists of two layers of sub-sumption structure (Section~\ref{controller}). The lowest $L$0.0 level is implemented in the body of each MR inside the beagle board which forms the main processor of each robot. The middle $L$0.1 level and $L$1 are implemented in desktop PCs where each level is separated for each MR. Thus, the entire two-layer control architecture is formed through the designated behaviors (Fig~\ref{masLayers}) for each MR in the system. 

The MRs carry out the task of pick-up or drop and carrying materials between different pair of ports. The $L$1 level controller in each MR is responsible for planning decisions to make them reach designated ports. The optimal path between different pair of ports is found out using Dijsktra's algorithm. These functions are carried out through the \textit{DECISION BEHAVIOR} (Fig~\ref{masLayers}) in $L$1 level. Dijsktra uses $X$($k$) as weight of an edge at every step of forming the path. $X$($k$) is estimated on real-time by Kalman Filtering at required $k$ in each MR. It was stated in Section~\ref{intro} that online estimation of $X$($k$) requires observation of the same at $k$-1. In this experiment, these observations are gathered from the beginning of first decision making and are used in subsequent calls for estimation. But, an MR may need to estimate the travel time of one or more edge which it did not traverse previously or has traversed long back. In this case, the observation of $X$($k$) for the concerned edge is not available. In this experiment, observation of $X$($k$) for the concerned edge at $k$-1 could not be obtained always and thus the available observation for the concerned edge is used. Thus, $X$($k$)s are estimated solely based on the historical observation of the concerned MR. Dijsktra's algorithm uses the estimated $X$($k$)s for each edge. Then it chooses the predecessor node of the current node, from which arrival to current node becomes least cost expending. This way the optimal path is formed using $X$($k$). These paths are shown in Section~\ref{res2}
\subsection{Results-I}
\label{res1}
\begin{figure*}[t]
\centering\includegraphics[scale = 0.28]{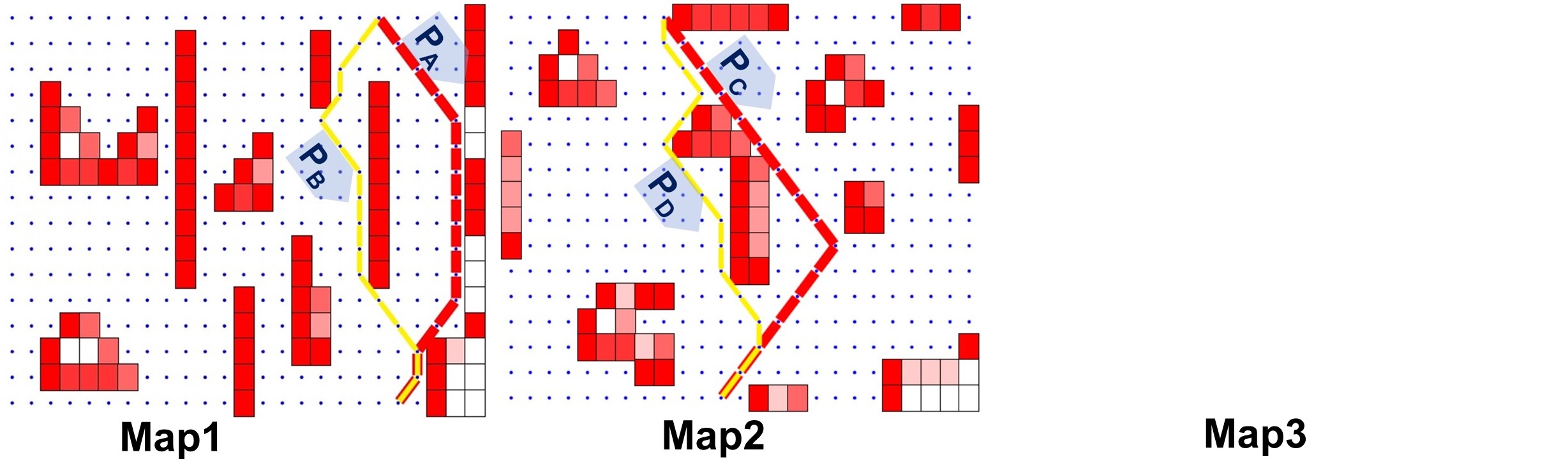}
\caption{change the pic, make it double column}
\label{outputPath}
\end{figure*}

The resultant paths (Figure~\ref{outputPath}) form a high level command or macro-command to be transferred to the next lower level $L$0.1. 

In $L$0.1, these high level commands or paths are disintegrated into macro-actions like \textbf{turn-left}, \textbf{turn-right}, \textbf{move-ahead}, \textbf{stop-left}, \textbf{stop-right}, \textit{et~cetera}. This disintegration happens due to the functions of three behaviors in $L$0.1 level (Fig~\ref{masLayers}).  
Further, these macro-actions are processed in $L$0.1 level to produce easy and low-level commands like \textbf{GO-$\langle$angle$\rangle$-$\langle$distance$\rangle$} which can be easily understood by the lowest level controller $L$0.0. These low-level commands enable the behaviors in $L$0.0 for accurate and prompt servo actions to generate movements in MR. This process is shown in Figure~\ref{pathByMR} where the MR traverses the path. Figure~\ref{pathByMR} shows different sections of the path in different steps \footnote{Video of path traversal is available at .....} from beginning (Part (A)) to end (Part (H)). 

In this way, the decentralized control is performed where each MR is the master of their own decision with the facility which is 
enabled due to the two layer subs-sumption control architecture based on behaviors. 
%This makes the whole MRS communicative and each MR help other MRs for their planning and control decisions.  of obtaining advice and help from others.
\subsection{Experiment-II} 
\label{exp2}
In this experiment, ontological data sharing is incorporated. The MRs are made to traverse repeatedly between different pairs of nodes. The pairs are designated previously from a list in order to suit the carriage necessity. The route computation between different pairs of node are done similarly as in Experiment~1 in Section~\ref{exp1}, using online estimated values of $X$($k$) as weight of edge. 

Online estimation of $X$($k$) at $k$ requires observation of the same at ($k$-1). In both Experiment~I and Experiment~II, these observations are gathered from the beginning of first decision making and are used in subsequent calls for path planning. However, when an MR needs to estimate the travel time of one or more edge(s) which it did not traverse previously, the available observation of $X$ for the concerned edge is at some previous instance which may not be ($k$-1) or close to that in many cases. These observations of distant past are used in experiment~I for estimating $X$($k$) at $k$. Thus, this will generate less accurate estimates. 

This is mitigated in this experiment~II by sharing the observation value from other MR who has travelled that edge in nearly previous instance. This way the observation of $X$ for the concerned edge at ($k$-1) or close will be available during estimation at $k$. 
The knowledge sharing contributes to estimate the travel cost for an unexplored edge at current instance in an MR. The behavior in $L$1 layer in each robot can ask the $L$1 level of other neighboring robots for observation values of $X$($k$) whenever required. $X$($k$)s are estimated for the necessary edges using observations either from the own MR or from neighbors.

Meanwhile, before deployment of the behavior-based system and ontology, some legacy data for travel times of different edges at different instances are obtained. These pool of date gathered by recording the travel times during the operation of MRS correspond to experiences of the MRs in the system. The estimated $X$($k$) values are compared to these legacy data to measure the accuracy which is discussed in Section~\ref{res2}. 

On the other hand, estimated $X$($k$) values of the relevant edges are used by Dijsktra's algorithm as weights of edges. These estimates are the main instrument at every step of deciding the predecessor to the current node.
Dijsktra's algorithm makes a node predecessor to current, when weight or cost from the former to later becomes minimum. Thus, accurate estimated value of $X$($k$) plays a vital role in deciding the predecessor to current node, in turn deciding the path. More accurate estimates contributes to generate paths with less total cost. Optimal paths are obtained with (experiment~II) and without (experiment~I) sharing the $X$($k$) values. These paths are compared in Section~\ref{res2}. 
\subsection{Results-II}
\label{res2}
This section tabulates the results of experiment~II. $X$($k$)s were estimated for each required edge at every step of Dijkstra's algorithm. These estimates are obtained using a non-linear model and
Kalman filtering, with observation data being shared from other MRs. These estimates are compared with that of experiment~I where estimation is done without sharing data among MRs. 
Figure~\ref{accuracy} plots the comparison of estimated $X$($k$) at different $k$ for different edges with that of legacy data obtained.

This section illustrates the comparison of paths and their costs obtained with and without sharing the travel times in the MRS. The path planning is done for 100 repetitions while increasing the $regression_no$ from 4 to 7.
\begin{figure}[h]
\centering\includegraphics[scale = 0.25]{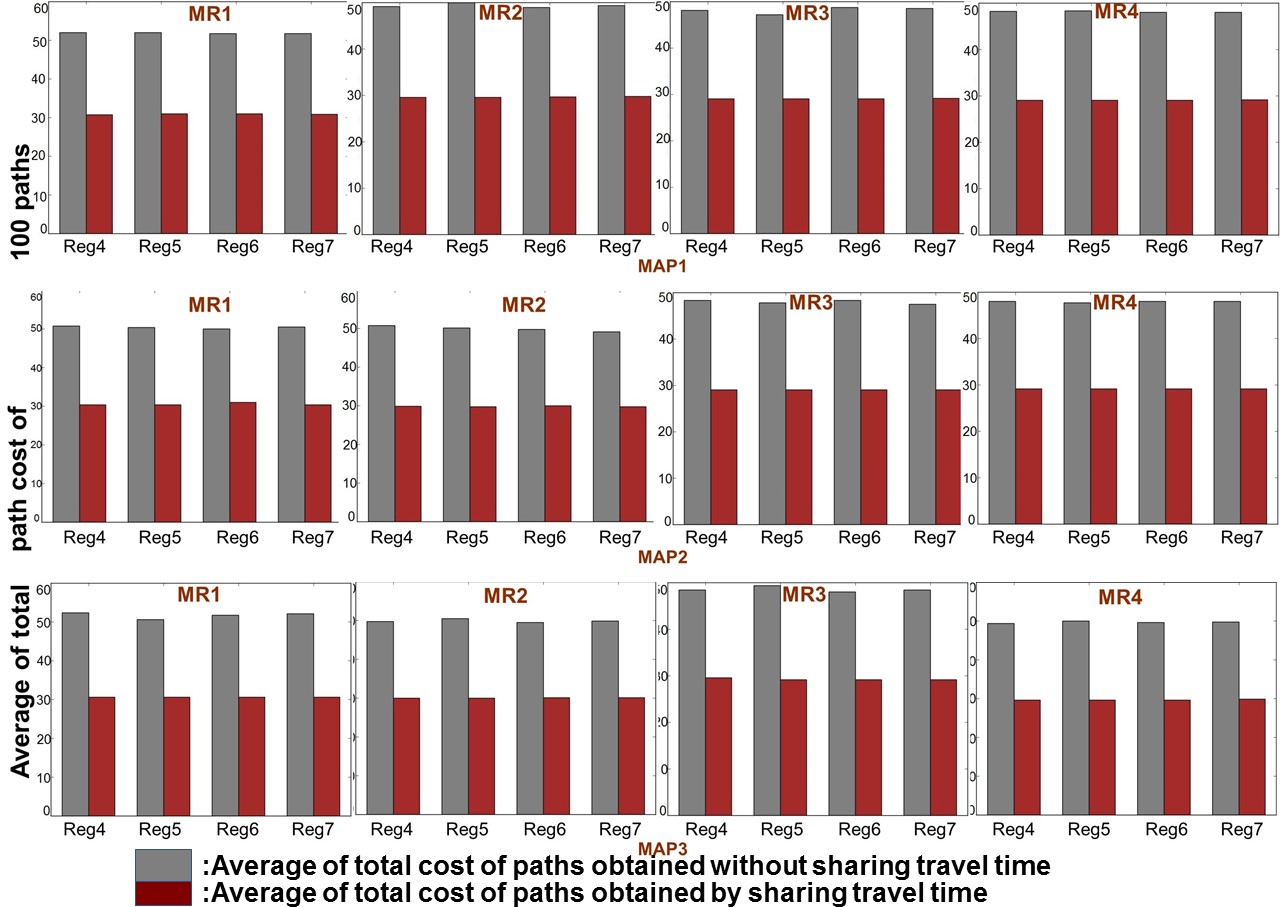}
\caption{Average of total path costs}
\label{pathCostsMap}
\end{figure}
Figure~\ref{pathCostsMap} illustrates average total path costs of 100 paths obtained in both Experiment~I and Experiment~II in four MRs operating in all three maps (Figure~\ref{3maps}). The average path costs of 100 paths obtained by sharing (Experiment~II) and  not sharing (Experiment~I) travel times are plotted for each regression, namely $Reg$4, $Reg$5, $Reg$6 and $Reg$7. 

For each regression in Map~1, the average of path costs obtained through collective intelligence are 40\% less than the average of path costs obtained without it. For each MR, the average of total path costs is almost same or vary in small margin with the increase of regression number. The reason of this is the lack of variation in environmental conditions. The travel times are varying on battery condition and floor. No other factor for affecting travel time could be incorporated in the laboratory set-up. 

On the other hand, the save of total path costs are same for all MRs in a single map. This signifies that paths found through collective intelligence in each MR is 40\% more cost efficient than the paths obtained without it. Thus, 
collective intelligence using travel time can affect to find more cost efficient paths in MRS. The average path costs decrease in case of Experiment~II as through collective intelligence more relevant observation of travel times are obtained in each MR. These values are instrumental for obtaining more accurate estimates of travel time. As a matter of fact, more accurate estimated values result in more optimal path with less cost than in that of obtained in Experiment~I. Few examples of these paths are discussed in the next section. 

Moreover, the save on total path costs is consistent in all the maps. Thus, the travel time is estimated better due to sharing of travel times from other MRs and this is true for all the representative structures of the floor. This signifies that more accurate estimation is possible through collective intelligence and this is independent of the structure of the floor. 
\subsection{Analysis of obtained paths}
This section illustrates few paths obtained in Experiment~I and Experiment~II under the same condition of regression no and MR.
\begin{figure}[h]
\centering\includegraphics[scale = 0.37]{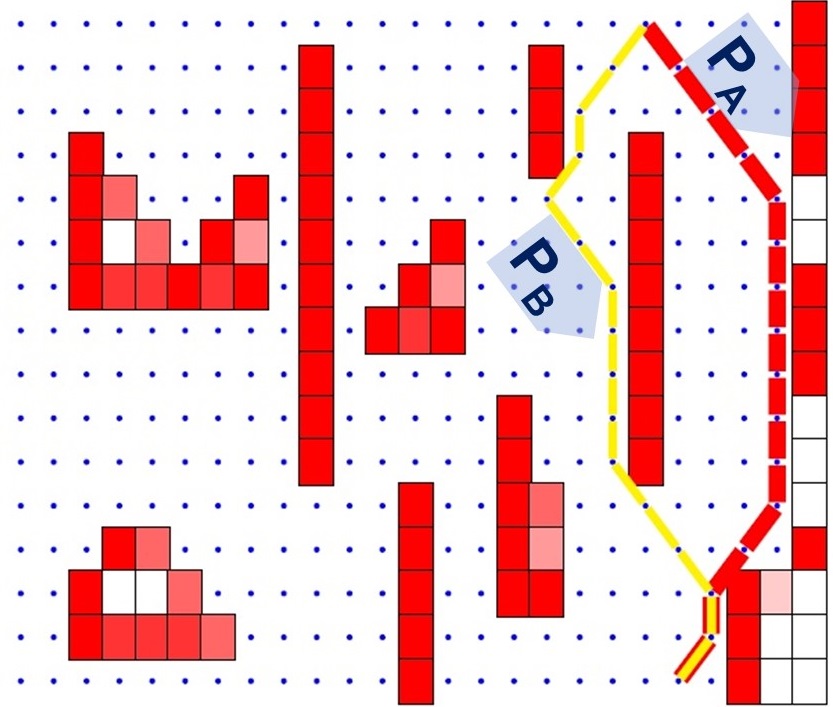}
\caption{Paths found by MR~1 in Map~1}
\label{pathA1M1}
\end{figure}
Figure~\ref{pathA1M1} plots two paths, $P_A$ and $P_B$ obtained in Map~1 for MR~1. $P_A$ and $P_B$ both have same source and destination. $P_A$ is obtained in Experiment~I in the third iteration of path planning, while $P_B$ is obtained in Experiment~II at the same iteration. Thus, they are both obtained at the same battery level and in the same map. Still, both the paths are different and have different total path cost.
As described in Section\ref{probContr}, $C_P$ denotes the cost of a path. $C_{PA}$ and $C_{PB}$ denote cost of $P_A$ and $P_B$ of Figure~\ref{pathA1M1} respectively.
The results show $C_{PA}$ = 66.5326 and $C_{PB}$ = 39.5385. Thus, 

\begin{align}
   C_{PB} < C_{PA}~by~40\% \nonumber
\end{align}

\begin{figure}[h]
\centering\includegraphics[scale = 0.37]{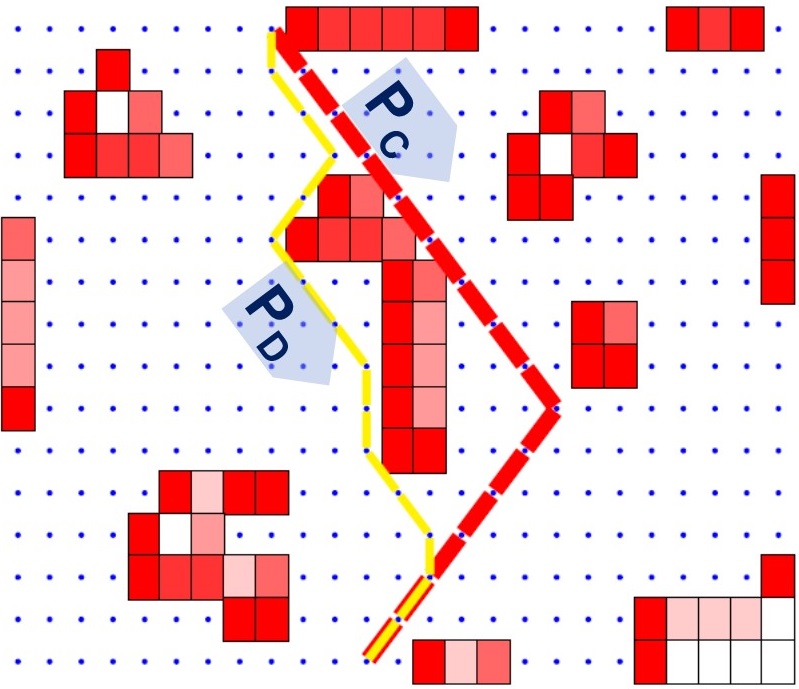}
\caption{Paths found by MR~2 in Map~2}
\label{pathA2M2}
\end{figure}
Figure~\ref{pathA2M2} plots two paths, $P_C$ and $P_D$ obtained in Map~2 for MR~2. $P_C$ and $P_D$ both have same source and destination. $P_C$ is obtained in Experiment~I in the third iteration of path planning, while $P_D$ is obtained in Experiment~II at the same iteration. Thus, they are both obtained at the same battery level and in the same map. Still, both the paths are different and have different total path cost.
$C_{PC}$ and $C_{PD}$ denote cost of $P_C$ and $P_D$ of Figure~\ref{pathA2M2} respectively.
The results show $C_{PC}$ = 58.0729 and $C_{PD}$ = 33.5707. Thus,

\begin{align}
   C_{PD} < C_{PC}~by~42\% \nonumber
\end{align}

From these two comparisons, it is evident that after sharing the travel times among the MRs, the path obtained in each 
MR have improved and are of less cost than that obtained without the sharing. 
%\subsection{Comparison with state of the art}
\section{Discussion and Conclusion}
\label{discConc}
The new method to compute cost parameter to be used in transportation and automation industry is proposed. With this new method, parameters now reflect the states of individual robots, their batteries and their environment. They usually arise locally at the robots as a result of performances of task.% It is interesting to mention here that route planning is done in Tesla's new X 75D model cars according to battery need. The route planner proposes breaks of variable times to enable recharging while travellers enjoy recess in driving. Thus, states of battery and environment are considered for optimal route planning in these models. 

%In context of automated logistics, the floor is designed in the form of a topology map. The robot carry out the task of traversing paths in the floor to different spots or ports. The travel time taken to traverse an edge in the map is identified as the key parameter. The experiments show that travel time varies with state of batteries and floor. Thus, travel time can reflect the condition of floor and batteries. 

In case of planning, the current state of robots and environment plays crucial role. The usual practice is to decide path using Euclidean distance and a path is considered optimal with optimal length or distance. Many industries (like BlueBotics \cite{Blue:2009}) use topology maps to describe the floor and employs a depth-first search to generates a length-optimal path. However, the true cost of traversing a path is not accounted in this case. The cost involved in traversing the path is generated from condition of floor, state of batteries, mechanical parts of robots. It is intuitive that an edge of same length will incur more cost in a rough floor than in smooth one. Thus, travel time is a better tool to decide a path than heuristics based on Euclidean distance. 

In this work, the decision making of each robot is based solely on the travel costs of its own.
In the dynamic estimation process, there are possibilities of not being able to learn observation of few edges due to lack of experience. Also, the observation gathered for a particular edge is too old to be relevant at the current instance of estimation. To address this, the sharing of travel time is incorporated to be able to share data of travel time from one MR to others. This enables the MR to generate more accurate estimation for travel times.

\bibliographystyle{plain}
\bibliography{elsarticle-template-1-num}
\end{document}